\documentclass[a4paper,11pt]{article}
\usepackage[T2A]{fontenc}			
\usepackage[utf8]{inputenc}			
\usepackage{graphicx}
\graphicspath{}
\DeclareGraphicsExtensions{.pdf,.png,.jpg}
\usepackage{amsmath,amsfonts,amssymb,amsthm,mathtools,verbatim,alltt} 
\usepackage{braket}
\usepackage{wasysym}
\usepackage{cancel}
\usepackage{setspace}
\usepackage{slashed}
\usepackage{cite} 
\usepackage[margin=1.5cm]{geometry}
\usepackage{authblk}
\everymath{\displaystyle}
\usepackage{mathrsfs}

\usepackage[colorlinks = true,
linkcolor = blue,
urlcolor  = blue,
citecolor = blue,
anchorcolor = blue]{hyperref}

\date{}
\title{Unruh effect in curved space-time and hydrodynamics}
\author[1,2,3]{R. V. Khakimov}
\author[1,2]{G. Yu. Prokhorov}
\author[1,2]{O. V. Teryaev}
\author[2,1,4]{V. I. Zakharov}
\affil[1]{Joint Institute for Nuclear Research, Joliot-Curie str. 6, Dubna 141980, Russia}
\affil[2]{NRC Kurchatov Institute, Moscow, Russia}
\affil[3]{Physics Department, Lomonosov Moscow State University, 1-2 Leninskie Gory, Moscow 119991, Russia}
\affil[4]{Pacific Quantum Center, 
Far Eastern Federal University, 10 Ajax Bay, Russky Island, Vladivostok 690950, Russia}

\begin{document}

\maketitle

\begin{center}
\section*{Abstract}
We consider an accelerated relativistic fluid in four-dimensional (anti-)de Sitter space-time. Analyzing only hydrodynamic equations, we construct the equilibrium stress-energy tensor.
We confirm that (A)dS vacuum corresponds to a thermal bath in the accelerated frame with a temperature, depending on the acceleration in a flat higher-dimensional (namely, five-dimensional) space, in which curved space-times are embedded. We develop the duality between hydrodynamics and gravity finding a direct relationship between the transport coefficients in flat and curved space-times.
\end{center}

\section{Introduction}

Hydrodynamics allows us to describe many physical systems and phenomena, not only classical, but also quantum. A particularly striking example is the extreme state of strongly coupled matter in the deconfinement phase, i.e. the quark-gluon plasma (QGP) found in heavy-ion collisions, which can be described as relativistic fluid system \cite{Schafer:2009dj}. The main tools of the hydrodynamic approach are the conservation laws (e.g., of the stress-energy tensor (SET) and currents), as well as the gradient expansion \cite{LandauFluid, Kovtun:2012rj}. The idea of the gradient expansion is that hydrodynamic quantities are given by series in terms of space-time derivatives acting on hydrodynamic degrees of freedom, such as temperature, chemical potential, fluid four-velocity ($T$, $\mu$, $u_{\mu}$). Considering hydrodynamics in curved space-time, we additionally take into account the gradients of the space-time metric $g_{\mu\nu}$. 
\\\\
A well-known example of the application of the first-order current gradient expansion is the work \cite{Son:2009tf} in which the relationship is established between seemingly very distant phenomena: hydrodynamics and gauge chiral anomaly (see also another original derivation \cite{Zakharov:2012vv}). This successful application of the methods of hydrodynamics to the analysis of quantum anomalies led to a series of subsequent works, such as \cite{Yang:2022ksq,Buzzegoli:2020ycf}, but mostly also for a gauge chiral anomaly. Therefore, of undoubted interest are the works on the study of another famous anomaly, the gravitational chiral anomaly \cite{Landsteiner:2011cp,Stone:2018zel,Landsteiner:2022wap,Frolov:2022rod}.
\\\\
This list includes recent work in the context of the hydrodynamic approach to anomalies \cite{Prokhorov:2022udo}, in which the relationship was established between the transport coefficients in flat space-time and the gravitational axial anomaly, which was called kinematical vortical effect (KVE). This effect is a manifestation of the duality between flat space-time and curved space-time (gravity), because it was shown that the axial current in a vortical and accelerated fluid in a flat space-time is determined by the gravitational chiral anomaly. In \cite{Prokhorov:2022udo} derivation of the KVE was made in Ricci-flat space-time approximation $R_{\mu\nu} = 0$ . The next logical step is to generalize our derivation and consider space-time with non-zero Ricci tensor, in the simplest case proportional to space-time metric $R_{\mu\nu} = \Lambda g_{\mu\nu}$ , so-called Einstein manifolds. This means that expressions for hydrodynamic quantities will contain terms with scalar curvature $R$. Finding the corresponding expansions in the case of the stress-energy tensor and the search for new elements of duality between hydrodynamics and gravity is the main goal of this paper.
\\\\
Besides duality, we will give a new perspective on the interesting question of the Unruh effect in curved space-time. It is known that the Hawking effect extends to other space-times with a horizon, in particular, there is a well-known analog of Hawking radiation in an accelerated frame, the Unruh effect \cite{Unruh:1976db, Volovik:2023stf}. The radiation temperature (the Unruh temperature) depends on the acceleration $T_U=\frac{|a|}{2 \pi}$, where $a_{\mu}=u^{\nu}\partial_{\nu}u_{\mu}$ is the proper acceleration, $|a|=\sqrt{-a^{\mu}a_{\mu}}$, and $ u_{\mu} $ is the four-velocity.
There is also another well-known example of the de Sitter space-time, whose temperature is determined by the scalar curvature $T_{R}=\frac{\sqrt{R/12}}{2\pi}$ \cite{Gibbons:1977mu}.
Later, the combined case with constant curvature and acceleration was considered, and the temperature of the radiation was shown to be a combination of $T_U$ and $T_R$ 
\begin{eqnarray}
T_{UR}= \frac{a_5}{2 \pi} = \sqrt{T^2_U + T^2_R} =\frac{\sqrt{|a|^2+R/12}}{2 \pi}\:.
\label{TDeser}
\end{eqnarray}
It is noteworthy that (\ref{TDeser}) is valid both for accelerated dS \cite{Narnhofer:1996zk, Deser:1997ri} ($R>0$) and anti-de Sitter (AdS) \cite{Deser:1997ri} ($R<0$) spaces. 
\\\\
This combined case became indicative, demonstrating the role of the flat higher-dimensional space-times in which curved space-times can be embedded. As in the Unruh effect, the temperature (\ref{TDeser}) is determined by acceleration, not four, but five-dimensional, the square of which is $|a_5|^2=|a|^2+R/12$. In this letter, we analyze the thermal radiation in accelerated (A)dS space from the point of view of relativistic hydrodynamics. We argue that (\ref{TDeser}) can be obtained from the basic hydrodynamic equations and the general relativistic covariance, if we know the hydrodynamic expansions in a curved space-time in the case with acceleration only. This approach develops the mentioned similarity between thermodynamics and gravity. 
\\\\
For clarity, let us start the consideration with the case of hydrodynamics in flat space-time and ordinary Unruh effect.

\section{Unruh effect from hydrodynamics: acceleration in flat space-time}
\label{sec2}

Let us consider a relativistic non-dissipative fluid with four-velocity $u_{\mu}$, constant acceleration $a_{\mu}$ and zero chemical potential $\mu = 0$ in flat space-time (we choose the signature $(+,-,-,-)$). For simplicity, the particles that form the fluid will be assumed to be massless (or nearly massless, when the mass is much less than the other dimensional parameters). The stress-energy tensor can be constructed in terms of gradient expansion that terminates at the fourth order \cite{Prokhorov:2019hif, Palermo:2021hlf, Ambrus:2021eod}.
The corresponding transport coefficients can be found within the quantum statistical approach from the correlators with boost operators. In particular, for the spins $s=0$ and $s =1/2$ we obtain \cite{Prokhorov:2019cik, Prokhorov:2019yft, Buzzegoli:2017cqy, Zakharov:2020ked}\footnote{Due to the covariance of (\ref{scalarSET}) and (\ref{spinorSET}), the metric can be both the Minkowski or, for example, the Rindler metric.}
\begin{eqnarray}
\langle \hat{T}^{\mu\nu} \rangle_{s=0} &=& \Big( \frac{4\pi^2 T^4}{90}  - \frac{|a|^4}{360\pi^2} \Big) u^{\mu} u^{\nu} - \Big( \frac{\pi^2 T^4}{90} - \frac{|a|^4}{1440\pi^2} \Big) g^{\mu\nu},
\label{scalarSET}\\
\langle \hat{T}^{\mu\nu} \rangle_{s=1/2} &=& \Big( \frac{28\pi^2 T^4}{180} + \frac{T^2|a|^2}{18} - \frac{17|a|^4}{720\pi^2} \Big) u^{\mu} u^{\nu} - \Big( \frac{7\pi^2 T^4}{180} + \frac{T^2|a|^2}{72} - \frac{17|a|^4}{2880\pi^2} \Big) g^{\mu\nu},
    \label{spinorSET}
\end{eqnarray}
It is essential that both SETs are equal to zero at the Unruh temperature \cite{Unruh:1976db}, which is a direct indication of the Unruh effect \cite{Becattini:2017ljh}
\begin{eqnarray}
  \langle \hat{T}^{\mu\nu} \rangle (T=T_U) = 0.
    \label{condition}
\end{eqnarray}
The explanation is the following. The matrix element of the SET is divergent and should be renormalized. A renormalization is used for which the matrix element for the Minkowski vacuum state (when there are no particles in the inertial frame) is zero. Relativistic covariance tells us that a tensor equal to zero in one reference frame is equal to zero in any other, including non-inertial ones \cite{Landau:1975pou}. Thus, it follows from the equality (\ref{condition}) that the finite temperature in the accelerated system corresponds to the Minkowski vacuum state. And this is, actually, the Unruh effect.\\\\
However, it is easy to show that a naive generalization of (\ref{scalarSET}) and (\ref{spinorSET}) to the case of curved space-time, with just a change of derivatives to covariant ones, e.g., $a_{\mu}=u^{\nu}\nabla_{\nu}u_{\mu}$, leads to non-conserved SETs. In particular (the details are given in the next Section), we would obtain from (\ref{scalarSET}) a non-zero contribution to the divergence in the case of the curved space-time, namely $\nabla_{\mu} \langle \hat{T}^{\mu\nu} \rangle_{(s=0)} = - \frac{a^2 R}{4320 \pi^2} a^{\nu} \neq 0$. This indicates that there are additional effects of curvature, that are not taken into account, which we consider in the next Section.

\section{Hydrodynamic gradient expansion: acceleration and constant curvature}

Consider now the same fluid, but in (A)dS space-time with $R_{\mu\nu}=\Lambda g_{\mu\nu}$, where $R=4\Lambda$ is a constant curvature. In this case, the gravitational field will be considered as external.
Gradient expansion should now also take into account the contribution of terms with scalar curvature $R=4\Lambda$ of the space-time (which is the second order in the gradients of the metric). Let us write down the expansion for the SET in the general case of massless particles with an arbitrary spin. Also, to facilitate the finding of covariant derivatives and further calculations, it will be convenient to switch to a dimensionless (or ``thermal'') acceleration $\alpha_{\mu} = \frac{a_{\mu}}{T}$. Therefore the SET in the fourth order in terms of gradients will have the form
\begin{eqnarray}
    \langle \hat{T}^{\mu\nu} \rangle &=& \Big( \rho_0 + A_1 \alpha^2 + A_2 R + B_1 \alpha^4 + B_2 \alpha^2 R + B_3 R^2 \Big) u^{\mu} u^{\nu} - \nonumber \\
    &&-\Big( p_0 + A_3 \alpha^2 + A_4 R + B_4 \alpha^4 + B_5 \alpha^2 R + B_6 R^2 \Big) g^{\mu\nu}+\nonumber \\
    && + \Big( A_5 + B_7 \alpha^2 + B_8 R \Big) \alpha^{\mu} \alpha^{\nu} + \mathcal{O}(\nabla^6),
    \label{curvedEMT}
\end{eqnarray}
where $\rho_0(T)$, $p_0(T)$, $A_i(T)$ and $B_i(T)$ are coefficients that depend on the only dimensional parameter, temperature $T$. We distinguished the coefficients corresponding to the terms of different orders of gradients: $A_i(T)$ correspond to the second order, and $B_i(T)$ to the fourth order. Some terms of expansion, e.g., $a^{\mu}u^{\nu}$, are dropped for reasons of PT-symmetries. Although, the expansion may include higher-order terms, which we denote as $\mathcal{O}(\nabla^6)$, but in most known similar cases (for example, systems with only acceleration (\ref{scalarSET}), (\ref{spinorSET})), the expansion of the SET of massless fields terminates at the fourth order \cite{Prokhorov:2019hif, Prokhorov:2019yft, Palermo:2021hlf}, i.e. $\mathcal{O}(\nabla^6)=0$. It is reasonable to assume that in the case under consideration there are also no higher-order terms.\\\\
The SETs (\ref{scalarSET}) and (\ref{spinorSET}) are now special cases of (\ref{curvedEMT}) in flat space-time limit. On the other hand, there is the well-known SET for the (A)dS vacuum state \cite{Page:1982fm, Dowker:1975tf}\footnote{The corresponding vacuum contribution is additional to the $\Lambda$-term, which is responsible for the geometric background.}
\begin{eqnarray}
  \langle \hat{T}^{\mu\nu} \rangle_{\text{vac}} = \frac{k}{4} R^{2} g^{\mu\nu}\,.
  \label{finalEMT001}
\end{eqnarray}
The conservation equations form the basis of hydrodynamics, in particular, for the SET, we have
\begin{eqnarray}
    \nabla_{\mu} \langle \hat{T}^{\mu\nu} \rangle = 0, \quad
   \langle \hat{T}^{\mu}_{\mu} \rangle = kR^2.
    \label{laws}
\end{eqnarray}
The second equation contains the famous gravitational conformal anomaly, and the numerical coefficient $k$ depends on the spin of the microscopic constituents \cite{Page:1982fm, Dowker:1976zf, Duff:1982yw}
\begin{eqnarray}
    \label{traceanomaly}
    k^{(s=0)} = \frac{1}{34560 \pi^2} \, , \quad
    k^{(s=1/2)} = \frac{11}{34560 \pi^2} \, .
\end{eqnarray}
We assume that system is in a global thermodynamic equilibrium,  which implies that the inverse temperature vector $\beta_{\mu} = \frac{u_{\mu}}{T}$ satisfies Killing equation \cite{Becattini:2016stj}
\begin{eqnarray}
 \nabla_{\mu} \beta_{\nu} + \nabla_{\nu} \beta_{\mu} = 0.
 \label{killingeq}
\end{eqnarray}
Chosen Killing vector is to be time-like and also future-oriented, therefore equilibrium condition possible for stationary space-times, as well as for static, as a special case. Note that for the static AdS space-time coordinates are global, while for the dS they cover only part of the manifold forming the static patch. The use of the condition (\ref{killingeq}) is dictated both by the physical formulation of the problem and by a significant simplification of the calculations, which actually allows us to study the higher order hydrodynamic effects in a gravitational field.
\\\\
To find the covariant derivative of the SET, we need to determine the covariant derivatives of temperature and acceleration. Using (\ref{killingeq}) and the definition of the Riemann tensor as a commutator of derivatives we obtain
\begin{eqnarray}
 \nabla_{\mu} \nabla_{\nu} \beta_{\alpha} = -R^{\rho}_{\:\:\mu\nu\alpha} \beta_{\rho}\,,
 \label{doubleder}
\end{eqnarray}
as for any Killing vector. Finally, using (\ref{killingeq}),(\ref{doubleder}) and the condition $u_{\mu} u^{\mu} = 1$ we obtain
\begin{equation}
    \begin{cases}
        \nabla_{\mu} T = T^{2} \alpha_{\mu}, \\
        \nabla_{\mu} \alpha_{\nu} = - T \alpha^{2} u_{\mu} u_{\nu} - \frac{R}{12\, T} (g_{\mu\nu} - u_{\mu} u_{\nu}).
    \end{cases}
    \label{derivatives}
\end{equation}
According to the first of the equations, the temperature gradient is the source of acceleration, which is consistent with the well-known Luttinger relation \cite{Luttinger:1964zz}\footnote{Thus, there are no external forces other than gravity and the SET is covariantly conserved in (\ref{laws}).}. To avoid unnecessarily cumbersome expressions, we will divide the tensor into two parts with different orders of gradients and consider them separately, which is possible, since each of the orders forms an independent system of equations.
\subsection{Second-order gradients}

The SET in the second order in gradients has the form
\begin{eqnarray}
  \langle \hat{T}^{\mu\nu} \rangle_{(2)} = \left( A_{1} \alpha^{2} +  A_{2} R \right) u^{\mu} u^{\nu} - \left( A_{3} \alpha^{2} +  A_{4} R \right) g^{\mu\nu} + A_5 \, \alpha^{\mu} \alpha^{\nu}.
  \label{secondEMT}
\end{eqnarray}
Combining derivatives from (\ref{derivatives}) we obtain
\begin{equation}
\begin{cases}
    \nabla_{\mu} A_i = \frac{\partial A_i}{\partial T} \nabla_{\mu} T = A'_i \, T^2 \alpha_{\mu}, \\
   \nabla_{\mu} \left( u^{\mu} u^{\nu} \right) = T \alpha^{\nu}, \\  
    \nabla_{\mu} \left( \alpha^{\mu} \alpha^{\nu} \right) = -\Big( T \alpha^{2} + \frac{R}{3T} \Big) \alpha^{\nu}, \\
    \nabla_{\mu} \alpha^2 = - \frac{R}{6T} \alpha_{\mu}.
\end{cases}
    \label{derivatives1}
\end{equation}
Then, the derivative of the SET will give us
\begin{eqnarray}
    \nabla_{\mu} \langle \hat{T}^{\mu\nu} \rangle_{(2)} = \Big[  (A_1 \alpha^2 T + A_2 RT)  - ( A'_3 \alpha^2 T^2 - A_3 \frac{R}{6T} + A'_4 RT^2) +(  -A_5 \alpha^2 T -A_5 \frac{R}{3T} + A'_5 \alpha^2 T^2 ) \Big] \alpha^{\nu}.
    \label{SETderivative}
\end{eqnarray}
Now we have to take into account that (\ref{SETderivative}) is zero and $\langle \hat{T}^{\mu}_{\mu} \rangle_{(2)} = 0 $, according to (\ref{laws}). Collecting terms of the same tensor structure and taking into account that (\ref{SETderivative}) are satisfied only when each independent term is zero, we obtain the following system of differential equations
\begin{equation}
\begin{cases}
    A_1  - A'_3 T - A_5 + A'_5 T = 0, \\
   A_2 T^2 + \frac{A_3}{6} - A'_4 T^3 - \frac{A_5}{3} = 0, \\
   A_1 - 4A_3 + A_5 = 0, \\
   A_2 -4A_4 = 0.
\end{cases}
    \label{diffsystem}
\end{equation}
Since the fields are massless and the only dimensional parameter is temperature, we know in advance the temperature dependence for $A_i(T)$
\begin{eqnarray}
    A_1 = \lambda_1 T^4, \quad A_2 = \lambda_2 T^2, \quad A_3 = \lambda_3 T^4, \quad A_4 = \lambda_4 T^2, \quad A_5 = \lambda_5 T^4.
\end{eqnarray}
where $\lambda_i$ are dimensionless constants. Therefore we transform (\ref{diffsystem}) into a system of algebraic equations
\begin{equation}
\begin{cases}
    \lambda_1 - 4 \lambda_3 + 3 \lambda_5 = 0, \\
   \lambda_2 + \frac{\lambda_3}{6} - 2 \lambda_4 - \frac{\lambda_5}{3} = 0, \\  
   \lambda_1 - 4 \lambda_3 +  \lambda_5 = 0, \\
    \lambda_2 -4 \lambda_4  = 0.
    \label{system1}
\end{cases}
\end{equation}
which can be easily solved. It follows that $\lambda_5 = 0$, $\lambda_2 = - \frac{\lambda_1}{12}$, $\lambda_3 = \frac{\lambda_1}{4}$, $\lambda_4 = -\frac{\lambda_1}{48}$ and finally (\ref{secondEMT}) has the form
\begin{eqnarray}
  \langle \hat{T}^{\mu\nu} \rangle_{(2)} = A \left( a^{2} - \frac{R}{12} \right) T^2 \Big( 4 u^{\mu} u^{\nu} - g^{\mu\nu} \Big),
  \label{secondEMT2}
\end{eqnarray}
where $A = \frac{\lambda_1}{4}$.
\\\\
Also, it is necessary to pay some attention to the zeroth order. Since in the zeroth order the trace anomaly is absent, simply from the conservation relations we obtain the standard expression
\begin{eqnarray}
  \langle \hat{T}^{\mu\nu} \rangle_{(0)} = \sigma T^4 \Big( 4 u^{\mu} u^{\nu} - g^{\mu\nu} \Big),
  \label{zeroEMT}
\end{eqnarray}
in accordance with the formulas (\ref{scalarSET}), (\ref{spinorSET}).
The coefficient $\sigma$ refers us to the Stephan-Boltzmann's law.

\subsection{Fourth-order gradients}

Now we write out the SET in the fourth order in gradients 
\begin{eqnarray}
  \langle \hat{T}^{\mu\nu} \rangle_{(4)} = \left( B_{1} \alpha^{4} + B_{2} \alpha^{2} R + B_{3} R^{2} \right) u^{\mu} u^{\nu}  - \left( B_4 \alpha^{4} + B_5 \alpha^{2} R + B_6 R^{2} \right) g^{\mu\nu} + \left( B_7 \alpha^{2} + B_8 R\right) \alpha^{\mu} \alpha^{\nu}.
  \label{fourthEMT}
\end{eqnarray}
Using (\ref{derivatives1}) we find the expression for the covariant derivative of the SET
\begin{eqnarray}
    \nabla_{\mu} \langle \hat{T}^{\mu\nu} \rangle_{(4)} &=& \Bigg[ \Big( B_1 \alpha^4 T + B_2 \alpha^2 RT + B_3 R^2T \Big) - \Big( B'_4 \alpha^4 T^2 - B_4 \alpha^2 \frac{R}{3T} + B'_5 \alpha^2 RT^2 - B_5 \frac{R^2}{6T} + B'_6 R^2T^2 \Big) + \nonumber \\
    &&+ \Big( B'_7 \alpha^4 T^2 - B_7 \alpha^4 T - B_7 \alpha^2 \frac{R}{2T} + B'_8 \alpha^2 RT^2 - B_8 \alpha^2 RT -B_8 \frac{R^2}{3T} \Big)\Bigg] \alpha^{\nu}.
    \label{fourthSET}
\end{eqnarray}
As in the previous section, substituting (\ref{fourthSET}) into (\ref{laws}) we have the system of differential equations
\begin{equation}
\begin{cases}
    B_2T^2  + \frac{B_4}{3} - B'_5T^3 - \frac{B_7}{2} + B'_8T^3 - B_8 T^2 = 0, \\
    B_1 - B'_4T + B'_7T - B_7 = 0, \\
   B_3T^2 + \frac{B_5}{6} -B'_6T^3 - \frac{B_8}{3} = 0, \\
   B_1 - 4B_4 + B_7 = 0, \\
   B_2 - 4B_5 + B_8 = 0, \\
   B_3 - 4B_6 = k.
\end{cases}
\end{equation}
but now it includes the trace anomaly. Again, we can move on to dimensionless constants
\begin{eqnarray}
    B_1 = b_1 T^4, \quad B_2 = b_2 T^2, \quad B_3 = b_3, \quad
    B_4 = b_4 T^4, \quad B_5 = b_5 T^2, \quad B_6 = b_6, \quad
    B_7 = b_7 T^4, \quad B_8 = b_8 T^2,\,\,
    \label{dimensionless}
\end{eqnarray}
where $b_i = \text{const}$, after that we are left with a system of algebraic equations
\begin{equation}
\begin{cases}
    b_2 =  - b_1/6 -b_8, \\
   b_3 = b_1/144 + b_8/3, \\ 
   b_4 = b_1/4, \\
   b_5 = b_2/4 + b_8/4 = - b_1/24, \\
   b_6 = b_3/4 - k/4 = b_1/576+ b_8/12 -k/4, \\
   b_7 = 0. 
   \label{system2}
\end{cases}
\end{equation}
Therefore (\ref{fourthEMT}) can be rewritten as
\begin{eqnarray}
  \langle \hat{T}^{\mu\nu} \rangle_{(4)} &=& B \left( a^{2} - \frac{R}{12} \right)^{2} \Big( 4 u^{\mu} u^{\nu} - g^{\mu\nu} \Big) + \frac{k}{4} R^{2} g^{\mu\nu} + b_8 \bigg[ \frac{R^2}{12} ( 4 u^{\mu} u^{\nu} - g^{\mu\nu}) + a^2 R ( \frac{a^{\mu}a^{\nu}}{a^2} - u^{\mu}u^{\nu}) \bigg],
  \label{fourthEMT1}
\end{eqnarray}
where $B = \frac{b_1}{4}$.
From the relativistic covariance, (\ref{fourthEMT1}) should transform to (\ref{finalEMT001}) in the vacuum state, which is possible for the term $R\, a^{\mu}a^{\nu}$ only when $b_8 = 0$, and therefore, finally we obtain
\begin{eqnarray}
  \langle \hat{T}^{\mu\nu} \rangle_{(4)} = B \left( a^{2} - \frac{R}{12} \right)^{2} \Big( 4 u^{\mu} u^{\nu} - g^{\mu\nu} \Big) + \frac{k}{4} R^{2} g^{\mu\nu}.
  \label{fourthEMT2}
\end{eqnarray}
Combining (\ref{secondEMT2}), (\ref{zeroEMT}) and (\ref{fourthEMT2}) together, we obtain the final full formula
\begin{eqnarray}
  \langle \hat{T}^{\mu\nu} \rangle = \Bigg[ \sigma T^4 + A \left( a^{2} - \frac{R}{12} \right) T^2 + B \left( a^{2} - \frac{R}{12} \right)^{2} \Bigg] \Big( 4 u^{\mu} u^{\nu} - g^{\mu\nu} \Big) + \frac{k}{4} R^{2} g^{\mu\nu}.
  \label{finalEMT}
\end{eqnarray}
Coefficients $A$ and $B$ for spins $s=0$ and $s=1/2$ can be defined from (\ref{scalarSET}), (\ref{spinorSET}). Taking into account equality $a^{\mu} a_{\mu} = - |a|^2$, for the case of $s=0$ expression (\ref{scalarSET}) will take the form 
\begin{eqnarray}
  \langle \hat{T}^{\mu\nu} \rangle_{s=0} = \Bigg[ \frac{\pi^2}{90 } T^4  - \frac{1}{1440 \pi^2} \left( |a|^{2} + \frac{R}{12} \right)^{2} \Bigg] \Big( 4 u^{\mu} u^{\nu} - g^{\mu\nu} \Big) + \frac{1}{960 \pi^2} \Big( \frac{R}{12} \Big)^2 g^{\mu\nu}\,.
  \label{scalarSET1}
\end{eqnarray}
And for the case of $s=1/2$ expression (\ref{spinorSET}) will change as 
\begin{eqnarray}
  \langle \hat{T}^{\mu\nu} \rangle_{s=1/2} = \Bigg[ \frac{7\pi^2}{180} T^4 + \frac{1}{72} \left( |a|^{2} + \frac{R}{12} \right) T^2 - \frac{17}{2880 \pi^2} \left( |a|^{2} + \frac{R}{12} \right)^{2} \Bigg] \Big( 4 u^{\mu} u^{\nu} - g^{\mu\nu} \Big) + \frac{11}{960 \pi^2} \Big( \frac{R}{12} \Big)^2 g^{\mu\nu}\,.\,\,\,\,\,\,\,
  \label{spinorSET1}
\end{eqnarray}
The formula (\ref{spinorSET1}) matches the result of \cite{Ambrus:2021eod} (that was obtained using quantum field theory in AdS space-time). 

\section{Discussion}

\subsection{Generalized Unruh effect: accelerated observer in (A)dS space-time}

Let us analyze the consequences of (\ref{finalEMT}). First, we should recall that in the limit $R\to 0$ the usual Unruh effect in flat space imposes the condition $\langle \hat{T}^{\mu\nu} \rangle (T = T_U) = 0$, as discussed in Section \ref{sec2}. This allows us to immediately obtain the connection between $\sigma, A$ and $B$, and (\ref{finalEMT}) takes the form
\begin{eqnarray}
  \langle \hat{T}^{\mu\nu} \rangle = \Bigg[ \sigma T^4 + A \left( a^{2} - \frac{R}{12} \right) T^2 + \left(\frac{A}{4 \pi^2}-\frac{\sigma}{16 \pi^4} \right) \left( a^{2} - \frac{R}{12} \right)^{2} \Bigg] \Big( 4 u^{\mu} u^{\nu} - g^{\mu\nu} \Big) + \frac{k}{4} R^{2} g^{\mu\nu}\,,
  \label{finalEMT00}
\end{eqnarray}
which, of course, both (\ref{scalarSET1}) and (\ref{spinorSET1}) satisfy. On the other hand, the vacuum SET should have the form (\ref{finalEMT001}). From the general relativistic covariance, the vacuum SET should have the form (\ref{finalEMT001}) in any reference frame, in particular, accelerated. Now we see that (\ref{finalEMT00}) at $T_{UR}$ leads to (\ref{finalEMT001})
\begin{eqnarray}
 \langle \hat{T}^{\mu\nu}\rangle(T=T_{UR})=\frac{k}{4} R^2 g^{\mu\nu}\,.
  \label{finalEMT02}
\end{eqnarray}
This means that the value of temperature $T_{UR}$ corresponds to vacuum, which therefore, is perceived by the accelerated observer as a heat bath with a temperature (\ref{TDeser}). Thus, the generalized Hawking-Unruh effect for accelerated observer in (A)dS space-time \cite{Narnhofer:1996zk, Deser:1997ri, Deser:1998xb} actually follows from the basic hydrodynamic equations (and the usual Unruh effect in flat space). This is a somewhat unexpected result, since our analysis essentially concerned only hydrodynamics in four-dimensional space-time, while (\ref{TDeser}) is associated with a five-dimensional acceleration $a_5$.

\subsection{Duality between hydrodynamics and gravity}

The result obtained add new elements to the duality between hydrodynamics and gravity 
previously discussed in \cite{Prokhorov:2022udo}. At first glance, these two approaches are essentially different - gravitational interaction is fundamental.  It is described, for example, by the vertex $\delta g_{\mu\nu} T^{\mu\nu}$ in action, where $\delta g_{\mu\nu}$ is the fluctuation of the space-time metrics. On the other hand, acceleration effects correspond to the macroscopic interaction $\alpha_{\mu}K^{\mu}$ in the density operator \cite{Becattini:2017ljh} with boost operator $K^{\mu}$. However, the analysis of conservation relations links the two types of effects. Indeed, for the SET (\ref{curvedEMT}), according to (\ref{finalEMT}) we obtain the relations
\begin{eqnarray}\nonumber 
&& A_1 = 4 A_3 = -12 A_2 T^2=-48 A_4 T^2\,, \\
&&B_1 = 4 B_4 =-6 B_2 T^2=-24 B_5 T^2 =144 B_3 T^4 = 576 B_6 T^4 +144 k T^4\,.
\label{dual1}
\end{eqnarray}
In this case,  $A_1$, $B_1$, $A_3$ and $B_4$ characterize the contribution, associated only with the acceleration $a_{\mu}$, which survives in the limit $R\to 0$, while $A_2$, $B_2$, $B_3$, $A_4$, $B_5$ and $B_6$ describe effects with the finite curvature $R$ (or mixed hydro-gravity effects). As follows from (\ref{dual1}), these two classes of effects are related to each other. 
\\\\
Since the hydrodynamic gradient expansion and conservation relations are quite general, our analysis is to be valid for any fluid with massless constituents with an arbitrary spin. However, at temperatures below the Unruh  temperature $T < T_U$ quantum phase transition occurs, a detailed analysis of which is given in \cite{Prokhorov:2023dfg}, and as the result (\ref{spinorSET}) changes. It should be expected that a similar constraint holds for (\ref{scalarSET1}), (\ref{spinorSET1}), and (\ref{finalEMT00}) when $T<T_{UR}$.

\section*{Conclusion}

We have considered a relativistic accelerated fluid of massless particles with an arbitrary spin in a thermodynamic equilibrium in a four-dimensional (A)dS space-time. We have derived the stress-energy tensor, which takes into account the effects of both acceleration and constant curvature.
\\\\
Conservation of the stress-energy tensor and general relativistic covariance allow us to find the linear equations relating the transport coefficients in flat and curved space-times. We have 
verified these relations directly in the particular case of the Dirac field in the AdS space-time. 
\\\\
The immediate consequence of the obtained formulas is a novel confirmation of the generalized Unruh effect for accelerated systems in (A)dS space-time: the temperature of the vacuum measured by the accelerated observer is determined by acceleration in a flat five-dimensional space-time.\\\\
{\bf Acknowledgements}
\\\\
The authors are thankful to D. V. Fursaev for stimulating discussions. 
The work of G.~Yu.~P. and V.~I.~Z. is supported by Russian Science Foundation Grant No. 22-22-00664.
The work of V. I. Z. is partially supported by Grant No. 0657-2020-0015 of the Ministry of Science and Higher Education of Russia.

\bibliography{lit}

\end{document}